\documentclass[12pt]{article}
\usepackage{graphicx}
\usepackage{amssymb}
\usepackage{amsmath}
\usepackage{url}
\usepackage{bm}

\newcommand\pubnumber{DPF2015-286}
\newcommand\pubdate{\today}
\newcommand{\vev}[1]{{\langle #1 \rangle}}
\newcommand{\bpsi}{{\bar \psi}}
\newcommand{\chib}{{\bar \chi}}

\newcommand{\nnn}{ \nonumber \\ }

\def\utah{${}^a$ Department of Physics and Astronomy \\
University of Utah, Salt Lake City, Utah, USA}
\def\frankfurt{${}^b$ Institut f{\"u}r Theoretische Physik - Johann Wolfgang Goethe-Universit{\"a}t \\
 Max-von-Laue-Str. 1, 60438 Frankfurt am Main, Germany}

\def\speaker{\footnote{Speaker}}

\def\Title#1{\begin{center} {\Large #1 } \end{center}}
\def\Author#1{\begin{center}{ \sc #1} \end{center}}
\def\Address#1{\begin{center}{ \it #1} \end{center}}
\def\andauth{\begin{center}{and} \end{center}}

\newcommand\pubblock{\rightline{\begin{tabular}{l} \pubnumber\\
         \pubdate  \end{tabular}}}
\newenvironment{Abstract}{\begin{quotation}  }{\end{quotation}}
\newenvironment{Presented}{\begin{quotation} \begin{center} 
             PRESENTED AT\end{center}\bigskip 
      \begin{center}\begin{large}}{\end{large}\end{center} \end{quotation}}
\def\Acknowledgments{\bigskip  \bigskip \begin{center} \begin{large}
             \bf ACKNOWLEDGMENTS \end{large}\end{center}}

\def\beq{\begin{equation}}
\def\eeq#1{\label{#1}\end{equation}}
\def\eeqn{\end{equation}}

\def\beqa{\begin{eqnarray}}
\def\eeqa#1{\label{#1}\end{eqnarray}}
\def\eeqan{\end{eqnarray}}

\let\bar=\overbar

\def\etal{{\it et al.}}

\def\vev#1{\langle #1 \rangle}

\def\Dslash{\not{\hbox{\kern-4pt $D$}}}
\def\dslash{\not{\hbox{\kern-2pt $\del$}}}

\def\msb{{\bar{\ssstyle M \kern -1pt S}}}

\begin{document}
\begin{titlepage}
\pubblock

\vfill
\Title{Magnetic Catalysis in Graphene}
\vfill
\Author{ Christopher Winterowd$^a$\speaker, Carleton DeTar$^a$,}
\Address{\utah}
\andauth
\Author{Savvas Zafeiropoulos$^b$}
\Address{\frankfurt}
\vfill
\begin{Abstract}
One of the most important developments in condensed matter physics in recent years has been the discovery 
and characterization of graphene. A two-dimensional layer of Carbon arranged in a hexagonal lattice, graphene exhibits 
many interesting electronic properties, most notably that the low energy excitations behave as massless Dirac fermions.
These excitations interact strongly via the Coulomb interaction and thus non-perturbative methods are necessary. Using
methods borrowed from lattice QCD, we study the graphene effective theory in the presence of an external magnetic field.
Graphene, along with other $(2+1)$-dimensional field theories, has been predicted to undergo spontaneous breaking of flavor symmetry including the formation of a gap as a result of 
the external magnetic field. This phenomenon is known as magnetic catalysis. Our study investigates magnetic catalysis using a fully non-perturbative approach. 
\end{Abstract}
\vfill
\begin{Presented}
DPF 2015\\
The Meeting of the American Physical Society\\
Division of Particles and Fields\\
Ann Arbor, Michigan, August 4--8, 2015\\
\end{Presented}
\vfill
\end{titlepage}
\def\thefootnote{\fnsymbol{footnote}}
\setcounter{footnote}{0}

\section{Introduction}
\vspace*{-.3cm}
The field of graphene has seen tremendous growth since its experimental verification in 2006 \cite{Novoselov}. The unusual electronic properties 
exhibited by monolayer graphene and its related structures have been partly responsible for driving this growth \cite{CastroNeto}. 

What makes graphene's electronic properties so fascinating is that the electrons interact strongly. The graphene fine-structure constant, $\alpha_g$, analogous to the fine-structure constant of QED, is of the order of one.
Although perturbative approaches have had success in elucidating aspects of graphene, one would also like 
a non-perturbative approach in order to investigate previous claims and explore new avenues in a controlled, systematic way. We do so here.

Magnetic catalysis, spontaneous symmetry breaking in the presence of a magnetic field, is thought to be a universal mechanism which is particularly applicable to field theories in $(2+1)$ dimensions \cite{Miransky1}. Using the low-energy effective field theory (EFT) describing graphene, studies applying the Schwinger-Dyson approach in the Hartree-Fock approximation find that a dynamical mass is generated.
The form of the dynamical mass in $(2+1)$ dimensions is given by $m_{dyn} \propto \alpha_g \sqrt{|eB|}$ \cite{Shovkovy}.

The low-energy EFT has an internal symmetry described by $U(4)$ which is broken down to $U(2) \otimes U(2)$ with the appearance of a nonzero value for the condensate $\vev{\bpsi \psi}$ and the generation of the dynamical mass. We use a discretized version of the EFT 
which employs staggered lattice fermions to correctly describe $N_f=2$ species of degenerate four-component Dirac spinors. Initial lattice studies did not see a nonvanishing condensate at arbitrary coupling \cite{Boyda}. Through careful attention to the zero-temperature and chiral limits, we show in a nonperturbative context that the graphene EFT does exhibit magnetic catalysis.
\vspace*{-.45cm}
\section{Graphene EFT}
\vspace*{-.3cm}
Graphene is predicted to be a semimetal from its band structure. One sees that around two inequivalent corners of the first Brillouin zone, called Dirac points, the dispersion is linear and the excitations are described by \cite{CastroNeto}
\beq
\mathcal{H}(\vec{k}) = \hbar v_F \vec{\sigma} \cdot \vec{k}.
\eeq{ValleyHamiltonian}
Counting the spin of the electron and the sublattice degree of freedom, one has a total of $8$ fermionic degrees of freedom which compose the $N_f=2$ species of four-component Dirac fermions in the EFT.
The quasiparticles interact via a Coulomb-type interaction described by 
\beq
 \mathcal{H}_{int} = \int d\vec{r} \int d\vec{r}~' \hat{\rho}\left(\vec{r}\right) \frac{e^2}{\epsilon | \vec{r} - \vec{r}~' | } \hat{\rho}\left(\vec{r}~'\right), 
\eeq{CoulombInteraction}
where $\hat{\rho}(r) = \Psi^{\dagger}(r) \Psi(r), ~\Psi^{\top}_{\mu} = \left( \psi_{\kappa A \mu}, \psi_{\kappa B \mu}, \psi_{\kappa' B \mu}, \psi_{\kappa' A \mu}\right)$ and $\epsilon$ refers to the dielectric constant of the substrate on which the graphene sheet is placed.
In the definition of $\Psi$, $\kappa$ and $\kappa'$ refer to the two Dirac points, $A$ and $B$ refer to the two inequivalent sublattices, and $\mu$ refers to the electron's spin. The effective coupling between quasiparticles is described by the graphene fine-structure constant which is given by $\alpha_g \equiv e^2/\epsilon v_F 4\pi > 1$.

\vspace*{-.45cm}
\section{Results}
\vspace*{-.3cm}
We employ one flavor of staggered fermions in $(2+1)$ dimensions. 
Interactions between the fermions are mediated by a scalar gauge potential which lives in $(3+1)$ dimensions. On the lattice this is described by the non-compact $U(1)$ Wilson action. For completeness we write our action below
\beqa 
S^{(NC)}_G &=& a^3_s a_t \frac{\beta}{2} \sum_n \sum^{3}_{i=1} \frac{1}{a^2_s}\left(\theta(n) - \theta(n+\hat{i})\right)^2, \\
S_F &=& a^2_s a_t \sum_{n} \bigg[ \frac{1}{2a_t} \chib_n \left(U_0(n)\chi_{n + \hat{0}} - U^{\dagger}_0(n-\hat{0})\chi_{n - \hat{0}}\right) \nnn
 &+& \frac{1}{2a_s}v_F\sum_{i=1,2} \eta_{i}(n) \chib_n \left(\chi_{n + \hat{i}} - \chi_{n - \hat{i}}\right) + m\chib_n\chi_n \bigg],
\eeqa{LatticeAction}
where $i$ labels the spatial coordinate and we have taken the spatial and temporal lattice spacings, $a_s$ and $a_t$, to be distinct. The mass term explicitly breaks the symmetry and is needed as an infrared regulator. We take the zero mass (chiral) limit to recover the EFT. The fermion action in (\ref{LatticeAction}) is known to have $O(a^2)$ splitting in the mass spectrum. To suppress this splitting at this order and obtain a better approximation to the continuum action, we 
replace the links $U_0(n)$ with tadpole-improved ``fat'' links and also add a third-nearest neighbor term. This results in what is known as the improved AsqTad action \cite{MILCStaggeredReview}.

\begin{figure}
\includegraphics[height=8cm,width=8cm]{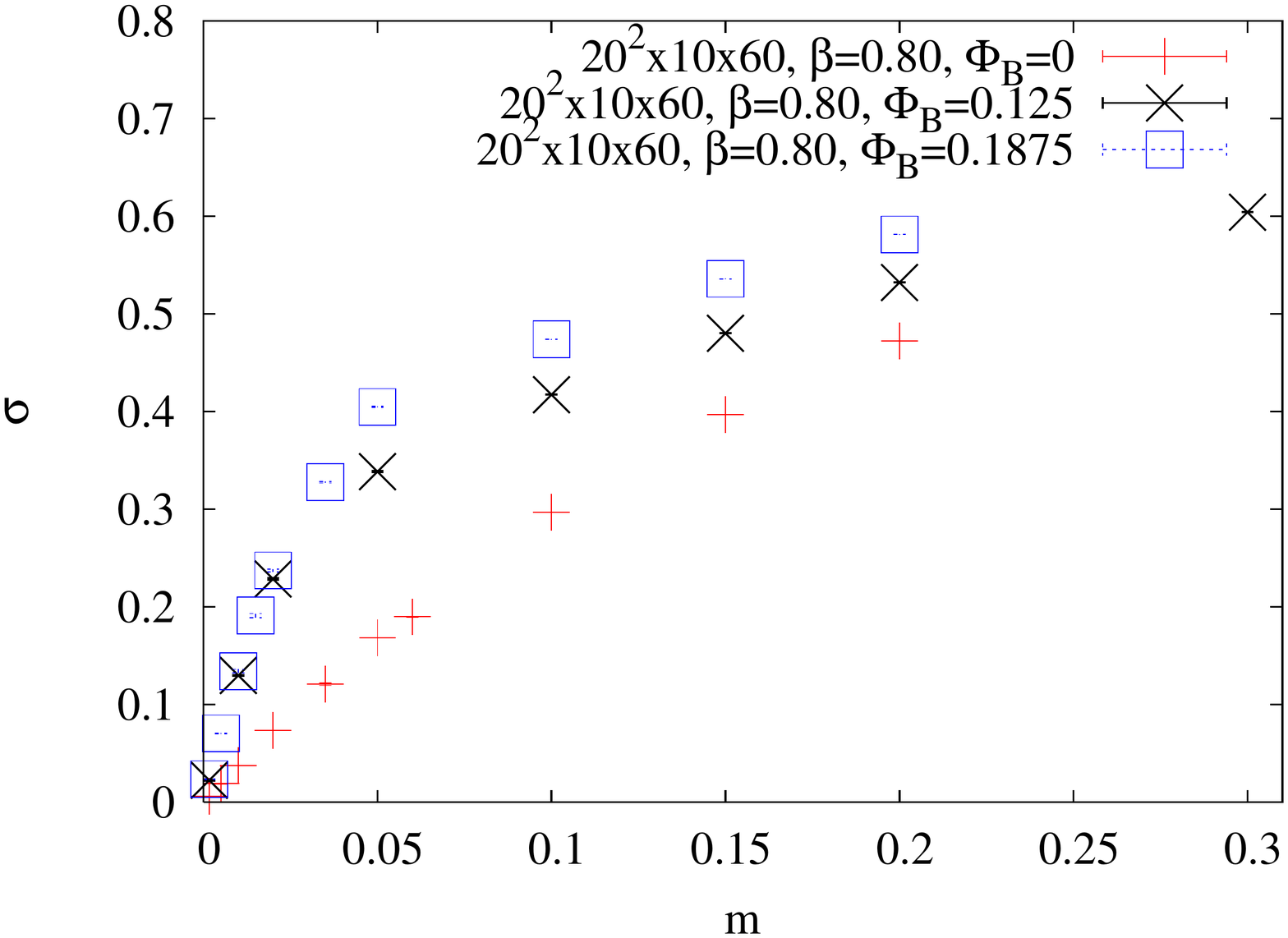} \hspace{-1cm}
\includegraphics[height=8cm,width=8cm]{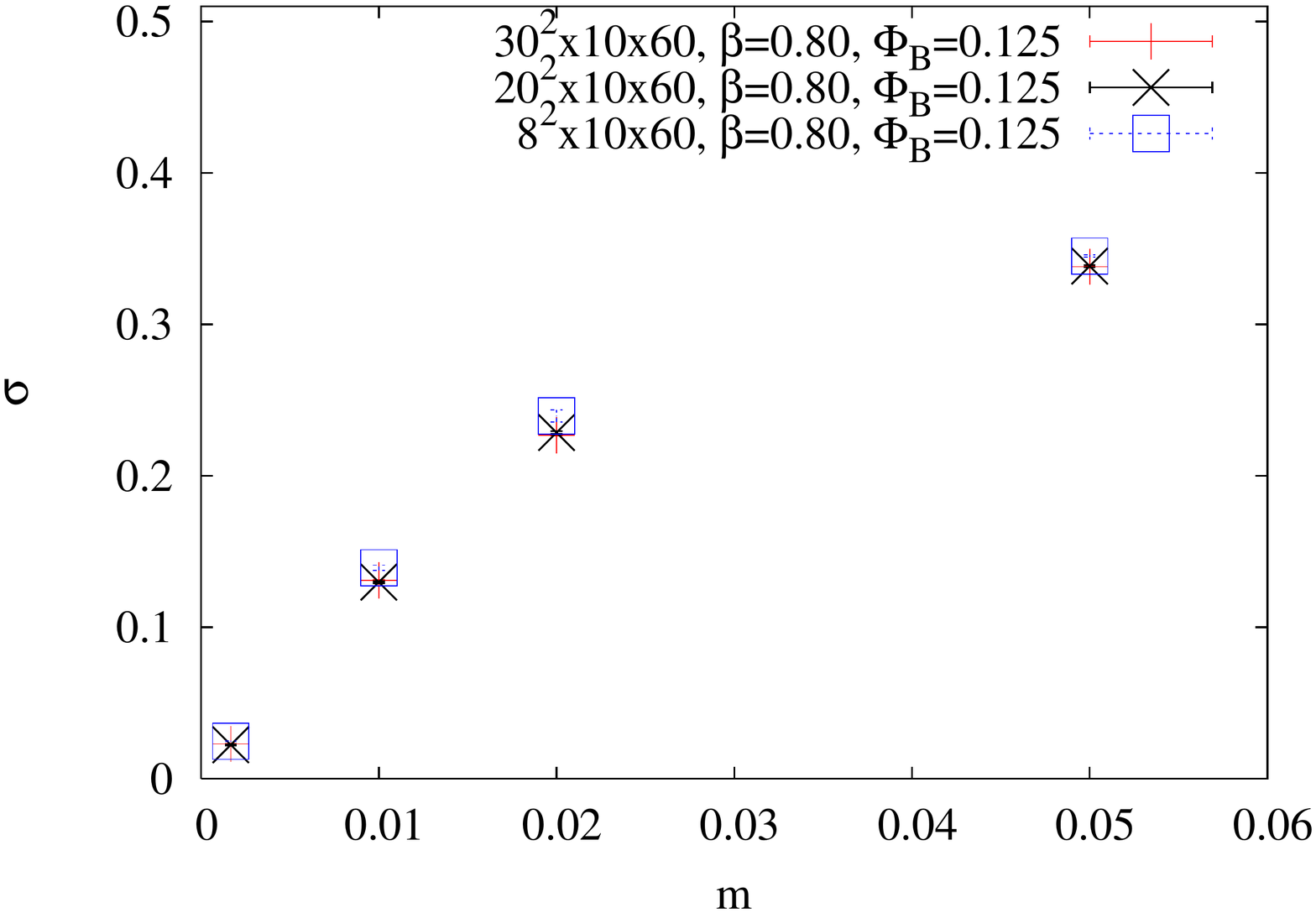}
\vspace*{-1.5cm}
\caption{(LHS) $\sigma \equiv \vev{\bpsi \psi}$ as a function of the bare fermion mass. We show the behavior at two values of the magnetic flux, $\Phi_B=0.125$ (black points) and $\Phi_B=0.1875$ (blue points) in units of $a^2_s$, as well
as at zero external magnetic field (red points). One notices that $\sigma$ vanishes with $m$ in the absence, as well as in the presence of, the external field. We argue that vanishing in the presence of an external field is a thermal effect. (RHS)  $\sigma$ as a function of mass with varying spatial volume $N^2_s$ for a magnetic flux $\Phi_B=0.125$. The lattice volumes are reported in the form $N_s^2\times N_z \times N_{\tau}$.}
\label{PBPComparison}
\end{figure}

It is known that the EFT undergoes spontaneous symmetry breaking at large coupling without an external magnetic field \cite{Drut}. It was thus necessary first to identify the symmetric phase as a function of the inverse coupling, $\beta \equiv 1/g^2$. 
Once the symmetric region was identified, we fixed the coupling within that region.
We then introduced the external magnetic field at several values of the quantized flux. The effect of the external field on the condensate can be seen in LHS Fig. (\ref{PBPComparison}). The extrapolation of $\sigma$ to zero as the bare mass is taken to zero signals that the symmetry is restored
when the explicit breaking term is removed. Typically, in lattice QCD, this is due to effects arising from the finite spatial size of the box. However, thermal effects can also play a role. The temperature, $T=1/N_ta_t$, is set by the temporal extent of the lattice. As one can see in RHS Fig. (\ref{PBPComparison}), the effects of a finite spatial volume are not significant 
and one still has $\sigma$ vanishing in the chiral limit. Further proof of this came from checking the screening masses of the lightest excitations of the theory.

\begin{figure} 
\includegraphics[height=8cm,width=8cm]{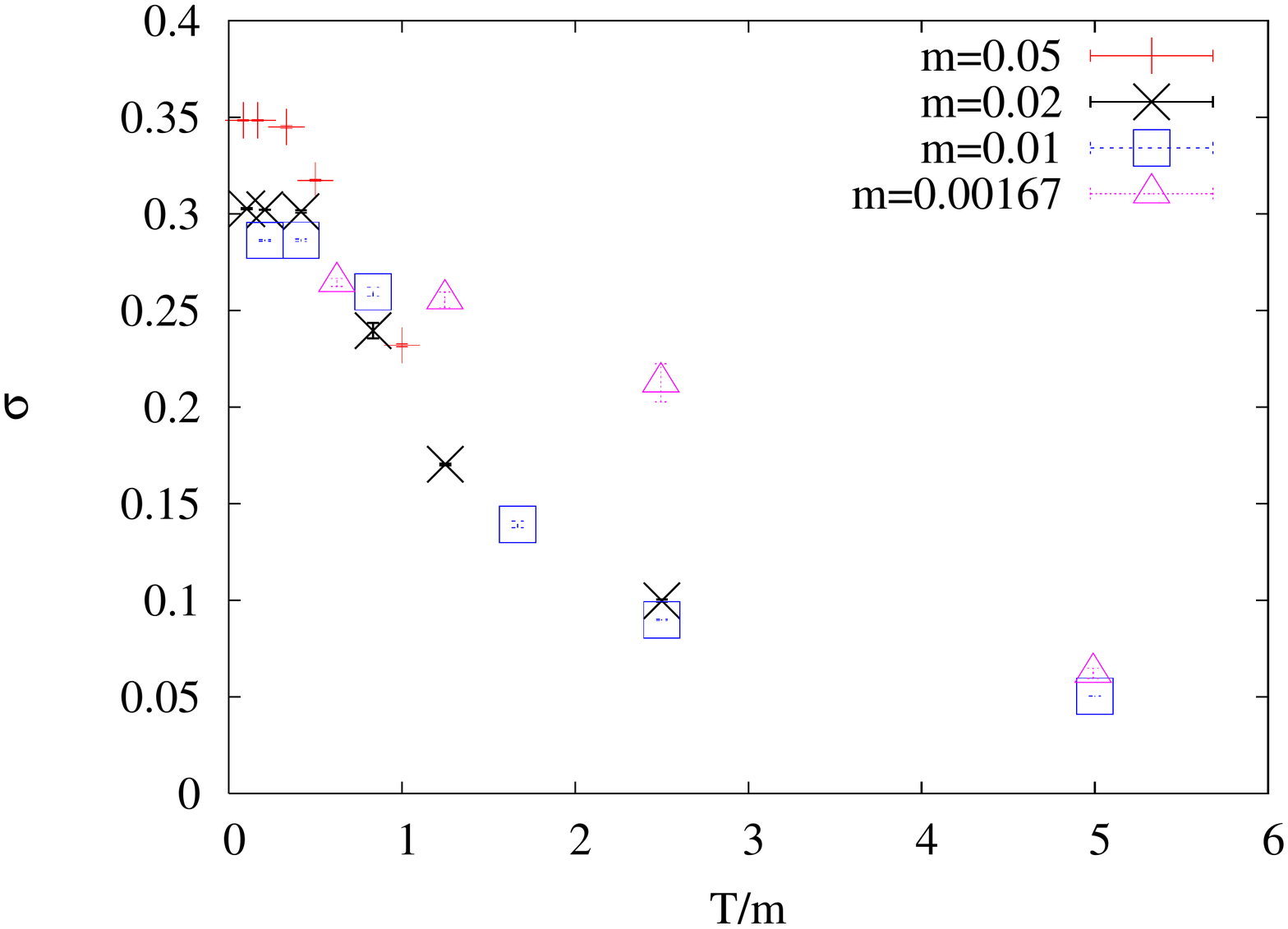} \hspace*{-1cm}
\includegraphics[height=8cm,width=8cm]{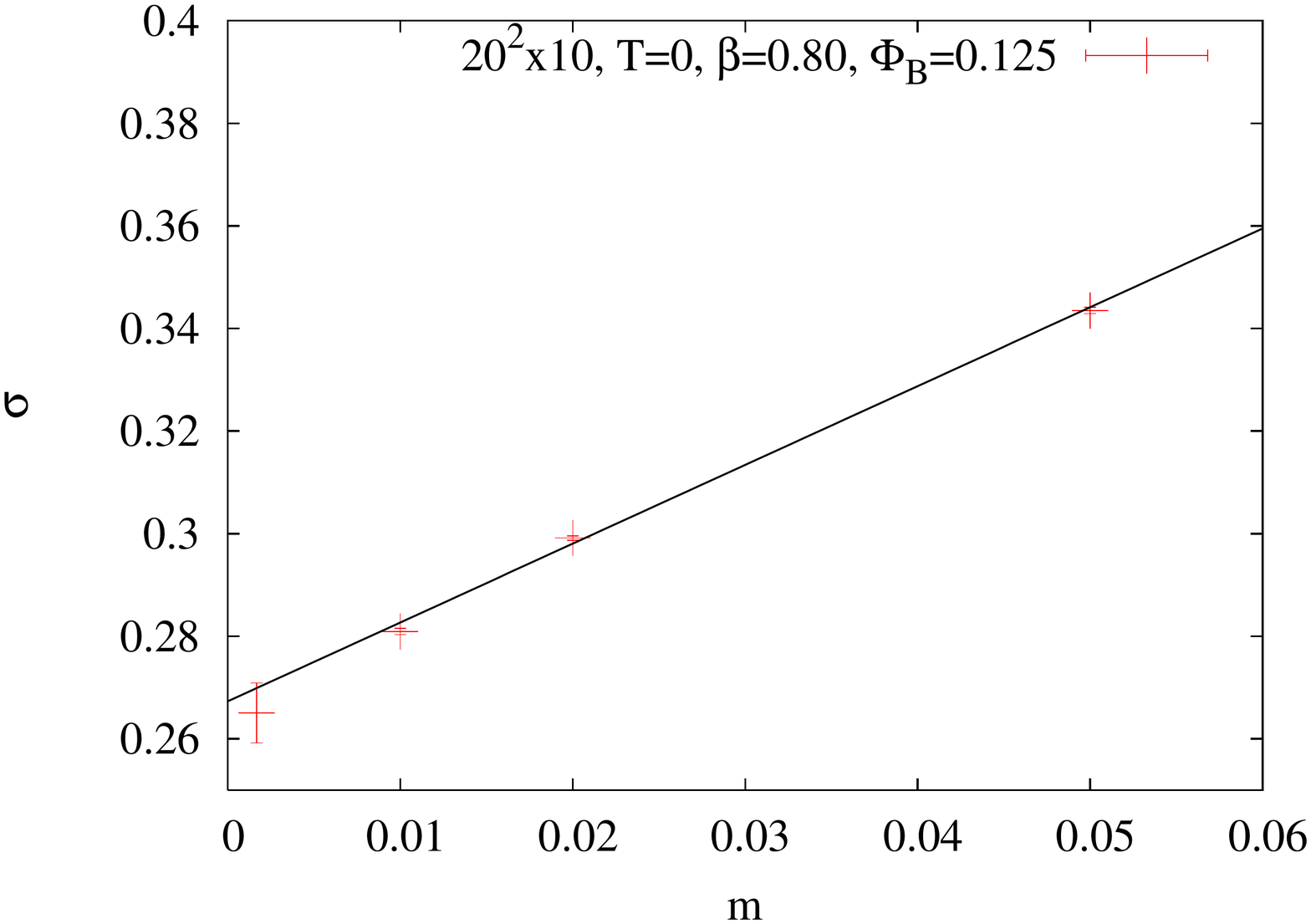}
\vspace*{-1.5cm}
\caption{(LHS) $\sigma$ plotted as a function of the scaling variable $T/m$. At small values of $T/m$, $\sigma$ tends toward a nonzero value while at large $T/m$ the condensate tends to zero. (RHS) Chiral extrapolation of $\sigma$ using the $T=0$ extrapolated points from the LHS. The nonzero intercept lends support to magnetic catalysis.}
\label{PBPvsT}
\end{figure}

From an analysis of $\sigma(T/m)$, illustrated in LHS Fig. (\ref{PBPvsT}), one can see that finite temperature effects strongly affect the chiral extrapolation. Namely, one sees that a
zero-temperature extrapolation must be performed before the chiral limit is taken as finite temperature effects become significant when $T/m \geq 1$. The zero-temperature extrapolated points are displayed in RHS Fig. (\ref{PBPvsT}), which lends support that a nonzero condensate exists in the chiral limit.

\vspace*{-.45cm}
\section{Conclusion and Outlook}
\vspace*{-.3cm}
The results of our fully non-perturbative calculation are consistent with magnetic catalysis. We have shown that finite volume and temperature effects are under control and that there is a systematic way of performing both
the chiral and zero-temperature limits. A nonzero value of the condensate was found at a single magnetic flux after performing both of the aforementioned limits.

Next, we plan to investigate the dependence of the condensate on the external magnetic flux and the coupling.
This will allow us to compare with the predictions of \cite{MiranskyGraphene}. 
\\
\vspace*{-.55cm}
\Acknowledgments
\vspace*{-.2cm}
This work was in part based on a variant of the MILC collaboration's public lattice gauge theory code. See \url{http://physics.utah.edu/~detar/milc.html}.
Calculations were performed at the Center for High Performance Computing at the University of Utah, Fermi National Accelerator Laboratory, and the LOEWE-CSC high performance
computer of Johann Wolfgang Goethe-University Frankfurt.
SZ would like to acknowledge the support of the Alexander von Humboldt foundation and fruitful discussions with Wolfgang Unger. CW and CD were supported by the US NSF grant PHY10-034278 and US DOE grant DOE DE-FC02-12ER41879.

\end{document}